\documentclass[twocolumn,superscriptaddress,aps,pre,bibtex]{revtex4-1}
\usepackage[T1]{fontenc}
\usepackage[latin9]{inputenc}

\usepackage[table,svgnames]{xcolor}
\usepackage{array}
\usepackage{units}
\usepackage{amsmath}
\usepackage{amssymb}
\usepackage{graphicx}

\begin{document}

\title{A simple unified view of branching process statistics: random
  walks in balanced logarithmic potentials}

\author{Serena di Santo} \affiliation{Departamento de
  Electromagnetismo y F{\'\i}sica de la Materia e Instituto Carlos I
  de F{\'\i}sica Te\'orica y Computacional. Universidad de Granada.
  E-18071, Granada, Spain} \affiliation{Dipartimento di Fisica e
  Scienza della Terra, Universit\`a di Parma, via G.P. Usberti, 7/A -
  43124, Parma, Italy} \affiliation{INFN, Gruppo Collegato di Parma,
  via G.P. Usberti, 7/A - 43124, Parma, Italy} \author{Pablo Villegas}
\affiliation{Departamento de Electromagnetismo y F{\'\i}sica de la
  Materia e Instituto Carlos I de F{\'\i}sica Te\'orica y
  Computacional. Universidad de Granada.  E-18071, Granada, Spain}
\author{Raffaella Burioni} \affiliation{Dipartimento di Fisica e
  Scienza della Terra, Universit\`a di Parma, via G.P. Usberti, 7/A -
  43124, Parma, Italy} \affiliation{INFN, Gruppo Collegato di Parma,
  via G.P. Usberti, 7/A - 43124, Parma, Italy} \author{Miguel
  A. Mu\~noz} \affiliation{Departamento de Electromagnetismo y
  F{\'\i}sica de la Materia e Instituto Carlos I de F{\'\i}sica
  Te\'orica y Computacional. Universidad de Granada.  E-18071,
  Granada, Spain}
\begin{abstract}
  We revisit the problem of deriving the mean-field values of
  avalanche exponents in systems with absorbing states. These are
  well-known to coincide with those of un-biased branching
  processes. Here, we show that for at least $4$ different
  universality classes (directed percolation, dynamical percolation,
  the voter model or compact directed percolation class, and the Manna
  class of stochastic sandpiles) this common result can be obtained by
  mapping the corresponding Langevin equations describing each of them
  into a random walker confined to the origin by a logarithmic
  potential.  We report on the emergence of non-universal
  continuously-varying exponent values stemming from the presence of
  small external driving --that might induce avalanche merging-- that,
  to the best of our knowledge, has not been noticed in the past.
  Many of the other results derived here appear in the literature as
  independently derived for individual universality classes or for the
  branching process itself.  Still, we believe that a simple and
  unified perspective as the one presented here can help (i) clarify
  the overall picture, (ii) underline the super-universality of the
  behavior as well as the dependence on external driving, and (iii)
  avoid the common existing confusion between un-biased branching
  processes (equivalent to a random walker in a balanced logarithmic
  potential) and standard (un-confined) random walkers.
\end{abstract}
\maketitle

Directed percolation (DP) is the paradigmatic example of a very large
class of systems --including catalytic reactions, growing interfaces
in random media, damage spreading, epidemic dynamics, and turbulence,
to name but a few--
exhibiting a phase transition separating a quiescent or absorbing
state from an active one
\cite{Liggett,Harris,Marro,Henkel,Odor,GGMA}. The essence of this very
robust universality class --which, curiously enough, had to wait long
for experimental backing \cite{Kaz}-- is parsimoniously encoded in the
following Langevin equation \cite{RFT1,RFT2,Henkel,Odor,GGMA}
\begin{equation}\dot{\rho}(\bold{r},t)= a
  \rho(\bold{r},t)-b\rho^2(\bold{r},t)+D\nabla^2\rho(\bold{r},t)+\sqrt{\rho(\bold{r},t)}
  \eta(\bold{r},t),
  \label{RFT} \end{equation}
where $\rho(\bold{r},t)$ is the density of activity at coordinates $\bold{r}$ and
time $t$, $a$ is the control parameter regulating the distance to the
critical point, $b$ and $D$ are constants, and
$\eta(t)$ is a 
Gaussian white noise of variance $\sigma^2$. Critical exponents,
scaling functions, and, in general, all critical features can be
obtained using Eq.(\ref{RFT}) as a starting point. The most
preponderant aspect of this equation, distinguishing it from other
classes, as for instance the Ising class \cite{Binney}, 
is the $\sqrt{\rho}$ factor in the noise amplitude. 
This square-root noise term stems from the ``demographic'' nature of
the particle-number  fluctuations; and it imposes that there are no 
fluctuations in the absence of activity, as corresponds to the absorbing
state \footnote{Another group of universal behavior is that of systems
  with noise proportional to the activity (rather that to the
  square-root of the activity); these encode a different type of
  processes where the most dominant fluctuations are not demographic,
  but associated to spatio-temporal variability in the overall
  parameters \cite{MN1,MN2,MN3}.}.

The same type of demographic noise also appears in other slightly
different universality classes, such as (i) the \emph{voter-model} or
neutral class describing the dynamics of neutral theories in which two
symmetric competing states are possible
\cite{Dickman,Liggett,Dornic01,AlHammal}; in this class there is no
deterministic force except for diffusion, and the noise amplitude is
different from zero only at the interfaces separating the two
absorbing states e.g. at $\rho=0$ and $\rho=1$,
i.e. $\dot{\rho}(\bold{r},t)=D\nabla^2\rho(\bold{r},t)+\sqrt{\rho(\bold{r},t)
  (1-\rho(\bold{r},t))}$ \cite{AlHammal}; (ii) the \emph{dynamical
  percolation} class \cite{DyP1,DyP2} --in which re-activation of
sites cannot occur and, as a consequence, the non-linear term in
Eq.(\ref{RFT}) needs to be replaced by a non-Markovian term $-\rho
(\bold{r},t) \int_{-\infty}^{t} dt' \rho(\bold{r},t')$ keeping track
of past activity while the noise term remains unchanged, and (iii) the
\emph{Manna} class of systems with many absorbing states such as
sandpiles in which an additional conservation law --that can be
encapsulated in an additional term $-\rho (\bold{r},t)
\int_{-\infty}^{t} dt \nabla^2\rho(\bold{r},t)$ \cite{FES-PRL,JABO1}--
exists, while the noise term remains as in directed percolation.

All systems with absorbing states, including these four classes and
some other more infrequent ones, not specified here-- share the common
feature of exhibiting avalanching behavior, meaning that if the
absorbing state is perturbed by a localized seed of activity, this can
trigger a cascade of events before falling back again into the
absorbing state. It is common knowledge that avalanches turn out to be
scale invariant at critical points; in particular, the avalanche-size
($S$) and avalanche-duration ($T$) probability distribution functions
can be written at criticality as
\begin{eqnarray}
P(S) &\sim& S^{-\tau} {\cal{G_S}}(S/S_C) \nonumber \\
F(T) &\sim & T^{-\alpha} {\cal{G_T}}(T/T_C),
\label{tau}
\end{eqnarray}
where ${\cal{G_S}}(S/S_C)$ and ${\cal{G_T}}(T/T_C)$ are cut-off
functions, and the cut-off scales, $S_C$ and $T_C$, depend only on
system size right at the critical point, and on the distance to
criticality away from it \cite{Leo}. Similarly, the averaged avalanche
size scales with the duration as $\langle S \rangle \sim T^\gamma$,
where the exponent $\gamma$ needs to obey the scaling relation
\cite{Sethna,Colaiori1},
\begin{equation}
\gamma=\frac{\alpha-1}{\tau-1}.
\label{scaling}
\end{equation}
In particular, for avalanches propagating in high dimensional systems
(or in densely connected networks) mean-field exponent values
$\tau=3/2$, $\alpha=2$ and $\gamma=2$ are obtained for all systems
with absorbing states. A compilation of avalanche exponents for
different dimensions and universality classes, as well as scaling
relationships, can be found in \cite{avalanches,Bona,Lubeck,Huynh}.

In order to explicitly compute these exponent values, textbooks
usually resort to the (Galton-Watson) branching process
\cite{Watson,Harris,Feller,Liggett}. In this, each node of a tree has
two branches emerging out of it; from an occupied/active node at
time/generation $n$ each of its two out-branches (at time/generation
$n+1$) are occupied/active with probability $p$ or left empty with
complementary $(1-p)$.  Observe that this is just a variant of
directed percolation running on a regular tree (see Figure 1).  For
illustration and completeness, we now present a very simple derivation
of its associated avalanche distribution functions.

To compute $P(S)$ --where $S$ is the total number of
occupied/active nodes before the process comes to its end-- one just
needs to evaluate the total number of connected trees of size $S$,
which is nothing but the Catalan number \cite{Catalan}
\begin{equation}
 C(S)=\frac{1}{S}  {{2S} \choose {S-1}},  
\end{equation}
and multiply it for the probability of each one to occur, $p^{S-1}
(1-p)^{S+1}$.  Evaluating the resulting expression $P(S,p) =
(2S)!/((S+1)! S!)  p^{S-1} (1-p)^{S+1}$ in the Stirling
approximation for $S>>1$, one readily obtains 
\begin{equation}
P(S,p)=\frac{\cal{N}}{\sqrt \pi} S^{-3/2} (4 p (1-p))^S,
\end{equation}
where $\cal{N}$ is a normalization constant; in particular, this
becomes a power law at the critical point $p=1/2$: $P(S,1/2)
=\frac{\cal{N}}{\sqrt \pi} S^{-3/2}$, implying $\tau=3/2$.  The
exponent $\gamma$ can also be derived using the statistics of branch
lengths in Catalan trees of a given size \cite{Ders}, leading readily
to the result $\gamma=2$; and from this, using the scaling relation
Eq.(\ref{scaling}), one obtains $\alpha=2$.

These results for the branching-process avalanche statistics can be
derived in a more systematic way --for different types of underlying
regular or random tree topologies-- within the generating function
formalism \cite{Redner,Plischke,SOBP}; indeed, already back in 1949
Otter computed the solution for the case of a Poissonian distribution
of branches per node \cite{Otter}.
\begin{figure}
\centering
\includegraphics[width=1.0\columnwidth]{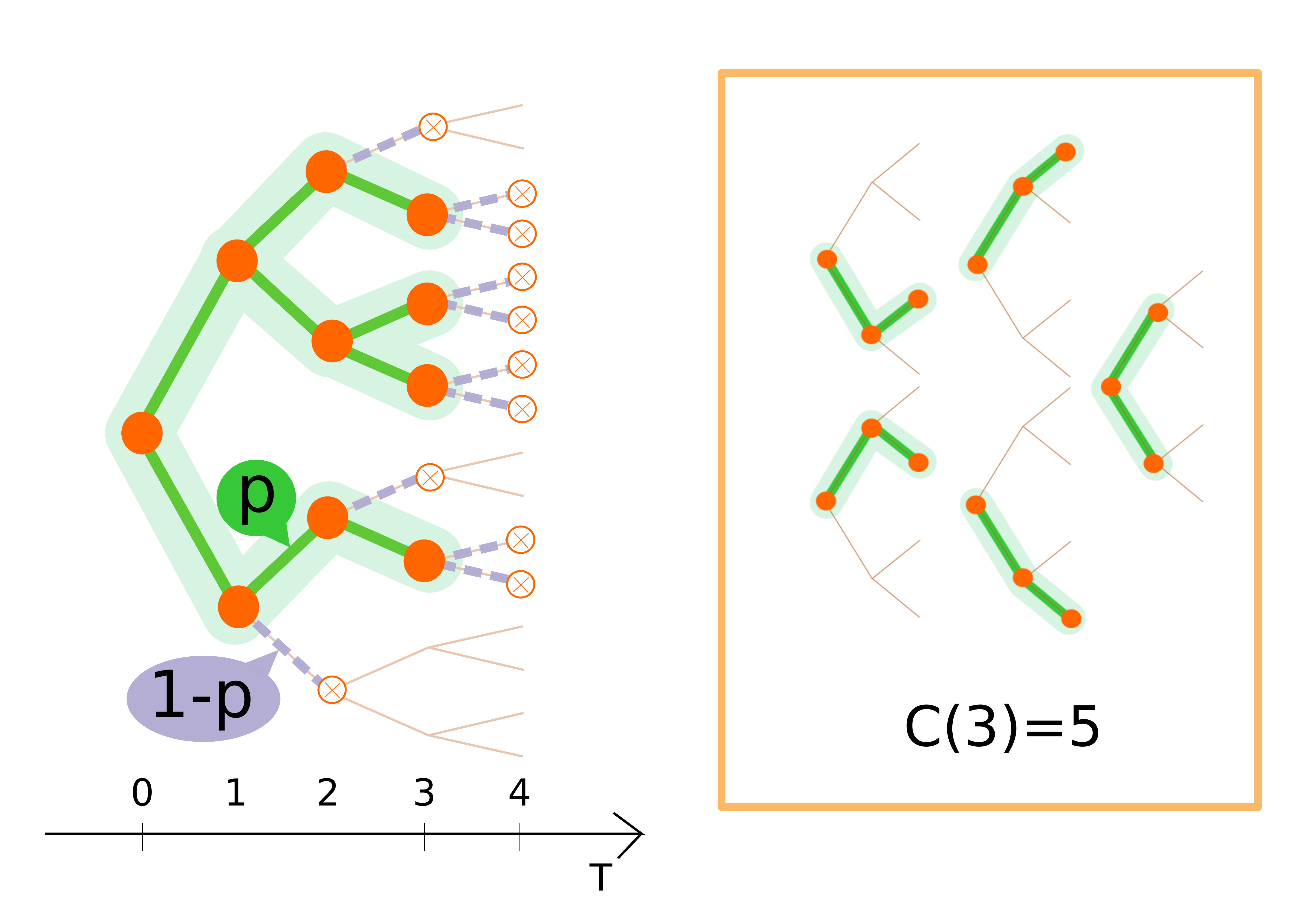}
\label{catalan}
\caption{Left: Illustration of a realization of the un-biased
  branching process, showing (highlighted) an avalanche of size $S=10$
  and duration $T=3$, together with the structure of the underlying
  rooted binary tree on top of which it unfolds.  Right: Visualization
  of the $5$ possible paths of $S=3$ as counted by the Catalan number
  $C(3)=5$. }
\end{figure}
Given that the result, e.g. a power-law with exponent $3/2$ for the
size distribution, is much more general than any specific branching
process in any specific tree-like topology, it is appealing from a
theoretical point of view to derive an even more general proof of
these results, covering all cases at once. From a slightly different
perspective, relying on field theory and scaling arguments
\cite{survival1,avalanches,survival2} the whole set of exponent values
can be obtained for each specific universality class, but again, the
result --being common to all classes, i.e. \emph{super-universal}--
should be amenable for a more generic explanation.

\begin{figure}[h]
\centering
\includegraphics[width=1.0\columnwidth]{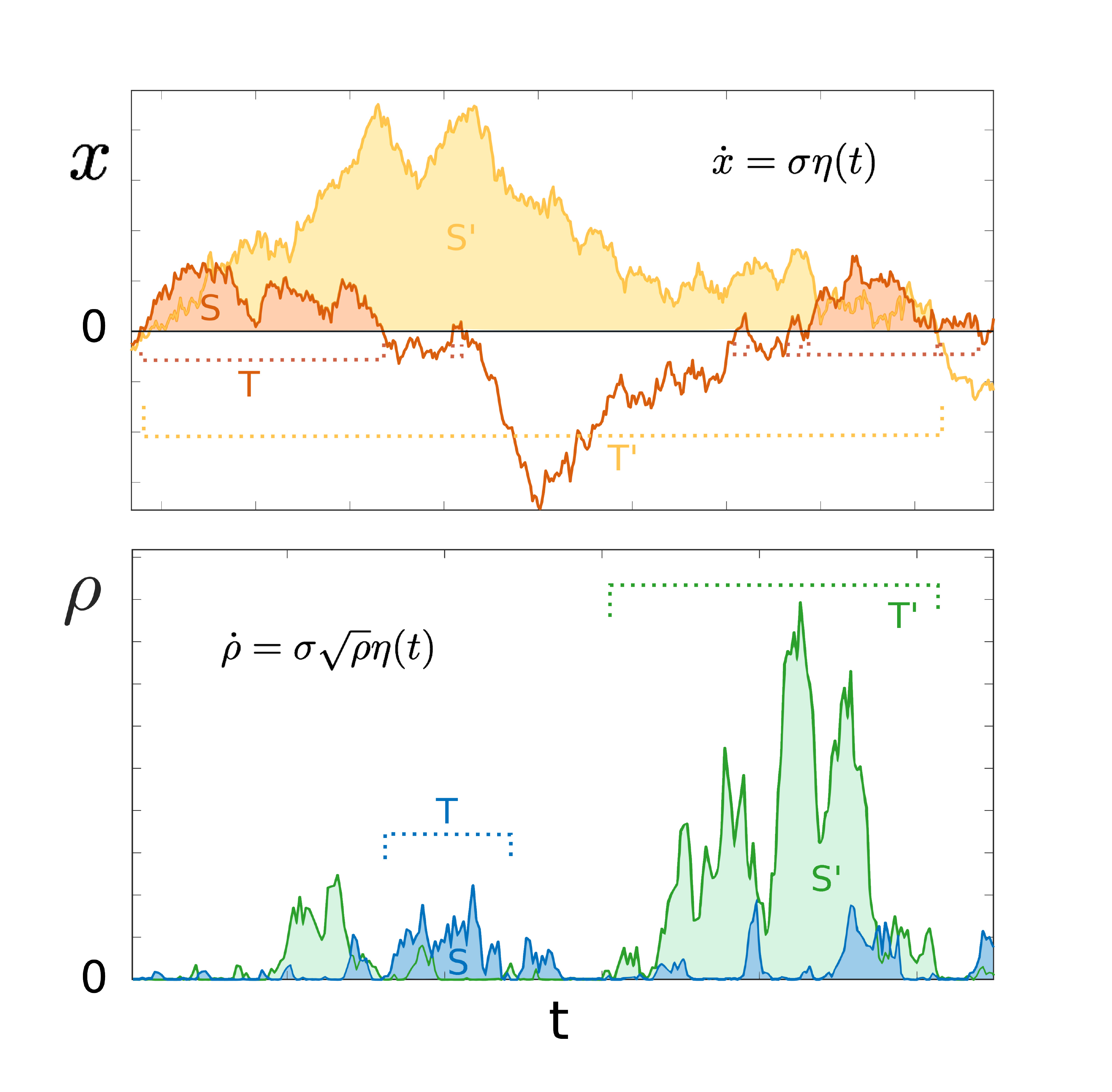}
\caption{Illustration of the time evolution of a standard random walk
  (RW) and a demographic random walk (DRW); each color corresponds to
  a different realization. Upper panel: standard RW that, in
  principle, can freely cross the origin. Avalanches start and end
  when the walker crosses the origin. Lower panel: the DRW can be
  represented as a stochastic RW moving in a balanced logarithmic
  potential that keeps the walker bounded to the origin. Since the
  variable is always strictly positive, the avalanches can be defined
  as the activity over a threshold $\epsilon\rightarrow0$.}
\label{sketch}
\end{figure}

The common feature shared by all the Langevin equations of the
different classes of systems with absorbing states, as already
mentioned above, is the presence of a demographic, square-root, noise
amplitude.  As a matter of fact --as illustrated in more detail in
Appendix A-- in the mean-field limit it is easy to derive a common and
unique effective Langevin equation for all classes of systems with
absorbing states at criticality, as
\begin{equation}
\dot{\rho}=\sqrt{\rho} ~\xi(t),
\label{1}
\end{equation}
where $\rho$ is the overall activity and $\xi(t)$ is a Gaussian white
noise with zero mean and $\langle\xi(t)\xi(t')\rangle=
2\sigma^{2}\delta(t-t')$ which needs to be interpreted in the It\^{o}
sense in order to guarantee that $\rho=0$ is an absorbing state
\cite{vankampen,Gardiner}. We refer to Eq.(\ref{1}) as ``demographic
random walker'' (DRW).  To avoid the complications of the It\^{o}
calculus, we write the equivalent equation in the Stratonovich
interpretation \cite{vankampen,Gardiner}:
\begin{equation}
\dot{\rho}=-\frac{\sigma^{2}}{2}+\sqrt\rho \eta(t)
\label{Strat}
\end{equation}
where now
$\langle\eta(t)\eta(t')\rangle=\frac{\sigma^{2}}{2}\delta(t-t')$.
Using now standard calculus to change variables to $x=\sqrt{\rho}$
directly gives \footnote{An alternative approach to analyze Langevin
  equations such as Eq.(\ref{1}) consists in reabsorbing the noise
  amplitude into the time-scale, leading to a standard random walk
  with a different ``clock'' \cite{Pruessner}. Another interesting
  possibility is deriving these results from a more general fractional
  Brownian motion \cite{Ding}.}
\begin{equation}
\dot{x}=-\frac{\sigma^{2}}{4x}+\eta(t).
\label{2}
\end{equation}
The resulting equation is just a
particular case of a one-dimensional random walker (RW) moving in a
logarithmic potential $U(x)=\lambda\log x$, i.e.
\begin{equation}
\frac{dx}{dt}=-\frac{dU(x)}{dx}+\eta(t)=-\frac{\lambda}{x}+\eta(t),
\label{3}
\end{equation}
where $\lambda$ is a positive constant and, in general,
$\langle\eta(t)\eta(t')\rangle=2\mu \delta(t-t')$, with $\mu$ a
generic positive constant.  Observe tha Eq(\ref{2}) corresponds to the
particular case, $\lambda=\mu=\sigma^{2}/4$ --that we call
\emph{balanced}-- in which the ratio between the amplitudes of the
logarithmic potential and the noise-correlation amplitude, $\mu$, is
equal to unity: $\beta \equiv \lambda/\mu=1$.  This perfect balance
between the deterministic-force and stochastic coefficients is
essential for what follows, as we shall see. More in general, let us
remark that, in the presence of an external field --allowing for the
spontaneous generation of activity at a fixed rate $h$-- Eq.(\ref{Strat})
needs to be complemented with an additional $+h$ term. Upon changing
variables, this implies $\beta=1 \rightarrow 1-h/\mu$, in Eq.(\ref{3})
and thus, in the presence of external driving, the perfect balance
between coefficients breaks down.

To compute avalanche exponents from Eq.(\ref{3}), let us define an
avalanche as a random walk $x(T)$, starting at $x(t=0)=0^+$ and
returning for the first time to the origin at time $T$, $x(T)=0$ (see
Figure 2).  The distribution or its return times is nothing but $F(T)$ as
defined in Eq.(\ref{tau}).  The problem of computing such a
return-time distribution for the random walk in a logarithmic
potential, i.e. by Eq.(\ref{2}), was solved by A. Bray \cite{Bray} and
revisited by F. Colaiori in the context of Barkhaussen crackling noise
\cite{Colaiori2}.  The solution requires writing down the equivalent
Fokker Planck equation for the Langevin dynamics, with a delta-like
initial condition centered at a value slightly larger than $x=0$, and
computing the probability flux $F$ at the origin as a function of the
time $T$ (more detailed sketch of the analysis is presented in
Appendix B for the sake of completeness). The resulting first-return
probability distribution function is
\begin{eqnarray}
  F(T)&=&\frac{4\mu\epsilon^{2\nu}}{\Gamma(\nu-1)}(1+\beta)(4\mu
  T)^{-\nu-1}e^{-\frac{x^{2}}{4\mu T}}  \nonumber \\
  &\sim&
  T^{-\nu-1}=T^{-\frac{3+\beta}{2}},
\label{return}
\end{eqnarray}
where $\nu=(1+\beta)/2$, implying $\alpha=\frac{3+\beta}{2}$.  Observe
that, in the limit of vanishing potential amplitude, $\lambda=0$, this
result reproduces the statistics of a freely-moving random walk, $F(T)
\sim T^{-\frac{3}{2}}$, while in the opposite perfectly-balanced
limit, $\lambda= \mu$ (i.e. $\beta=1$) the result is $F(T) \sim
T^{-2}$ in agreement with the expectations for the un-biased branching
process.  It is noteworthy that --despite the fact that the random
walk in a logarithmic potential gives a non-universal avalanche
duration exponent-- for the undriven DRW case, in which the
logarithmic potential derives from a change of variables in It\^{o}
calculus, there exists a perfect balance between the coefficients of
the equation; they both depend on the noise amplitude and,
compensating each other, they generate the universal value
$\alpha=2$. However, as said above, in the presence of an external
field, $\beta = 1-h/\mu$ breaking down the perfect balance between
coefficients, non-universal continuously-varying avalanche exponents
appear (see Figure 3); in particular,
 \begin{equation} \alpha= 2-\frac{h}{2\mu}.
\end{equation}  

In any possible discrete/particle model with absorbing states, this
change of exponents stems from the fact that --owing to the external
driving-- avalanches from different initial seeds (each of them
spontaneously generated by the external driving field) can merge,
which allows their combination to survive longer and be larger, thus
leading to smaller effective exponents $\alpha$ and $\tau$ (see Table
1).

\begin{figure}
\centering
\includegraphics[width=1.0\columnwidth]{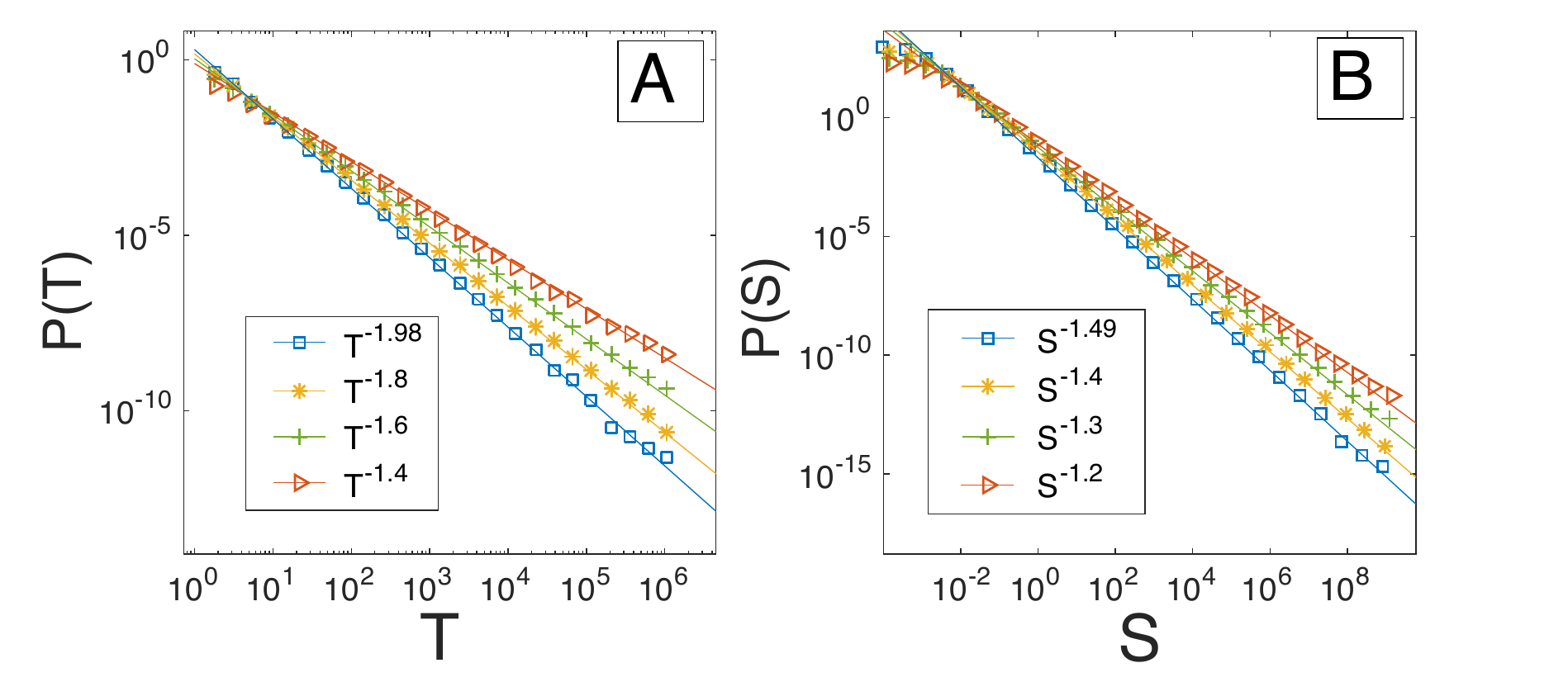}
\caption{Size-avalanche and duration-avalanche distributions for the
  un-driven demographic random walk as described by Eq.(\ref{1}), as
  well as for diverse values of the external driving field (marked
  with symbols) $h=0.01$ (blue squares), $h=0.1$ (yellow stars),
  $h=0.2$ (green crosses) and $h=0.3$ (red triangles), with reference
  curves (solid lines) $t^{-2+h/(2\mu)}$ and $s^{-3/2+h/(4\mu)}$ (as
  derived in the text), respectively, illustrating the agreement with
  theoretical predictions.}
\end{figure}

Turning back to the general discussion, using the above result
together with simple scaling, we can readily derive the associated
avalanche size exponent, $\tau$.  In order to have a unified notation
let us use a generic variable $v(t)$, which can be in particular,
$x(t)$ for the RW, or $\rho(t)$ for the DRW. The size of any given
avalanche is defined as the area under the curve defined by the random
walk, i.e.  $S=\int_0^{T} v(t) ~ dt$, and we are interested in the
distribution of such sizes as a function of $T$, $P(S|T)$.  Given that
the typical displacement of a random walk in time $t$ scales as $v
\sim \sqrt{t}$, for the DRW (for which there is an additional
square-root factor) we have $v \sim \sqrt{v} \sqrt{t}$, and thus, $v
\sim t$; hence, we can write, in general, $v\sim t^{\phi}$, with
$\phi=1/2$ and $\phi=1$ for the RW and the DRW (either driven or
undriven), respectively.

It is natural to define a new rescaled variable
$\tilde{v}(t/T)=v(t)/T^{\phi}$ which describes a random excursion in
the interval $[0,1]$.  In these terms,
\begin{equation}
  S=\int_0^{T} v(t) dt ~\sim ~T^{\phi+1}\int_0^{1} \tilde{v}(z)dz.
\label{rescaling}
\end{equation}
Thus, the average avalanche size, $\langle S\rangle$ obtained
averaging over all possible avalanche shapes, $\tilde{v}(z)$, scales
also with $T^{\phi+1}$, implying $\gamma=\phi+1$.  

Using the previous result, $P(S|T)$ can be written as a scaling form
$P(S|T) = T^{-\gamma} {\cal{G}}\left(S/T^{\gamma} \right)$ where the
factor $T^{-\gamma}$ comes from the normalization condition,
and the unspecified scaling function ${\cal{G}}$ obeys
${\cal{G}}(z)\ge 0$ for all $z$ and $\int_0^{\infty} {\cal{G}}(z) dz=1$.
Having computed the conditional probability $P(S|T)$, we can
explicitly obtain $P(S)$ as
\begin{eqnarray}
P(S)&=& \int_0^{\infty} dT P(S|T) F(T) \nonumber \\
&\sim& C \int _0^{\infty} dT ~ T^{-\gamma} T^{-\alpha}
{\cal{G}}\left(S/T^{\gamma} \right)     \nonumber \\
&\sim& C S^{-(\gamma+\alpha-1)/\gamma} \int _0^{\infty} du  u^{\frac{(\alpha-1)}{\gamma}} {\cal{G}}\left(u\right),
\end{eqnarray} 
and, thus, $\tau=(\gamma+\alpha-1)/\gamma$ (which is nothing but the
scaling relation Eq.(\ref{scaling})). Plugging the value of
$\alpha$ and $\gamma$ derived above one obtains the well-known result
$\tau=4/3$ for the standard random walk \footnote{In the case of the
  standard RW case the scaling function ${\cal{G_{RW}}}$ has been
  exactly derived (see e.g. \cite{Satya}), but its specific form is
  not essential for our purposes here.} and 
\begin{equation}
\tau=\frac{3}{2}-\frac{h}{4\mu},
\end{equation}
for the DRW, which reduces to the well-known result $\tau=3/2$ for the
un-driven case. Table 1 contains a summary of the exponents for the
different cases.

\begin{table}
  \setlength{\tabcolsep}{0pt}
\begin{tabular}{m{0.25\linewidth}>{\centering}m{0.22\linewidth}>{\centering}m{0.22\linewidth}>{\centering}m{0.31\linewidth}}
\cline{2-4} 
 & Unbiased RW & Demographic RW & Driven demographic RW\tabularnewline
\hline
\rowcolor{FloralWhite}
\vspace{2mm}
$P(T)\sim T^{-\alpha}$\vspace{2mm}
 & {\Large{}$\ensuremath{\alpha=\nicefrac{3}{2}}$} & {\Large{}$\ensuremath{\alpha=2}$} & {\Large{}$\ensuremath{\alpha=2-\nicefrac{h}{2\mu}}$}\tabularnewline
\hline 
\vspace{2mm}
$\ensuremath{P(S)\sim S^{-\tau}}$\vspace{2mm}
 & {\Large{}$\ensuremath{\tau=\nicefrac{4}{3}}$} & {\Large{}$\ensuremath{\tau=\nicefrac{3}{2}}$} & {\Large{}$\ensuremath{\tau=\nicefrac{3}{2}-\nicefrac{h}{4\mu}}$}\tabularnewline
\hline 
\rowcolor{FloralWhite}\vspace{2mm}
$\ensuremath{P(S|T)\sim T^{-\gamma}}$\vspace{2mm}
 & {\Large{}$\ensuremath{\gamma=\nicefrac{3}{2}}$} & {\Large{}$\ensuremath{\gamma=2}$} & {\Large{}$\ensuremath{\gamma=2}$}\tabularnewline
\hline 
\end{tabular}
\caption{Summary of the avalanche exponents for the standard RW, for
  the demographic RW, and for the driven demographic RW (in
  the  presence of an external field,
  allowing for the spontaneous generation of activity at a fixed rate $h$).}
\end{table}

Results beyond critical exponents have also been obtained in the
literature, for example, the average shape of random-walk excursions
is a semi-circle for standard un-biased random walkers
\cite{Colaiori1} while it is a parabola for demographic walkers
\cite{Zapperi}. This can be easily seen by rescaling the walks to
$\tilde{v}$ and the times to $t/T$ to collapse curves as described
above.  In this way $\tilde{v}(t/T) = {\cal{F}}(t/T)$ where
${\cal{F}}(t/T)$ is a scaling function.  Given that, $v(t)\sim
t^{\gamma-1}$, dividing by $T^{\gamma-1}$, $\tilde{v}(t/T) \sim
(t/T)^{\gamma-1}$, at least for small times, $t<<T$. Considering that
a similar relation holds for the reverse time walk starting from
$t/T=1$, then the avalanche shape is ${\cal{F}}(t/T)=[(t/T)
(1-t/T)]^{\gamma-1}$ which is a semicircle for $\gamma=3/2$ (RW) and a
parabola for $\gamma=2$ (DRW and driven DRW).

In summary, we have explicitly shown that the mean-field values of
avalanche exponents in systems with absorbing states can be computed
in a general way by mapping them into a random walk confined by a
logarithmic potential, Eq.(\ref{2}).  Of course, this same conclusion
could have been reached by arguing in a heuristic way that all of
high-dimensional processes involving absorbing states should be
effectively described by an un-biased branching process, and then
constructing a continuous description of it (i.e. a Fokker-Planck or
equivalently a Langevin equation) which would be nothing but
Eq.(\ref{1}).

An interesting corollary is that the exponents do change in the
presence of spontaneous creation of activity, even if the rate is
arbitrarily small. This result, which stems from the marginality of
the associated logarithmic potential could be relevant to understand
empirical results; for instance in cortical networks, avalanches of
neural activity have been reported to exhibit branching process
statistics \cite{BP2003}; still inspection of some of the
most careful estimations reveals possible deviations from $\tau=3/2$
\cite{Schuster}, which could be potentially ascribable to a
non-vanishing inherent spontaneous-activation.

We hope that this short paper will help avoiding in the future the
frequent confusion we have encountered (mostly in the neuroscience
literature) about branching processes and their relation with random
walks and also in interpreting empirical results considering the
possibility of non-universal continuously-varying exponents.

\appendix

\section{Irrelevance of non-linear terms}
For the directed percolation class in the mean-field limit, where
spatial heterogeneity is neglected, Eq.(\ref{RFT}) reduces to
\begin{equation}
\dot{\rho}(t)= a \rho -b \rho^2 +\sqrt{\rho} \eta(t).
\label{MF}
\end{equation}
At criticality, i.e.  $a=0$, there is still a non-linear (saturation)
term $-b\rho^2$ which introduces a characteristic maximal activity
scale, thus apparently precluding scale-invariance.  The way out of
this apparent conundrum is that when studying avalanches in
discrete/particle models, activity is created at a single location,
and in the continuous limit, this corresponds to vanishing density of
activity, $\rho=0$.  Thus, one needs to consider a large but finite
system size, say $\Omega$ (e.g. one could think of a fully connected
network with $\Omega$ nodes), and perform a finite-size scaling
analysis. Defining $y$ by $\rho=y/\Omega$ then --up to leading order
in $\Omega$-- Eq.(\ref{MF}) reduces to $\dot{y}(\tilde{t})=\sqrt{y}
\eta(\tilde{t})$ where $\tilde{t}=\Omega t$. In other words, employing
the correct rescaled variables $y$ and $\tilde{t}$ the saturation term
is never ``seen'' by the expanding avalanche, which is compatible with
the density being equal to zero, as the avalanche invades an
infinitely large system. Observe that in the main text we keep the
notation with $\rho$ and $t$, for the sake of simplicity.

Similarly, the voter-model (or compact directed percolation
\cite{compact} or neutral theory) class --characterized by two
symmetric absorbing states-- is described, as said above, by the
Langevin equation \cite{AlHammal}
\begin{equation} 
  \dot{\rho}(t) = D \nabla^2 \rho(\bold{r},t)+\sqrt{\rho(1-\rho}) \eta(\bold{r},t),
  \label{rho} \end{equation}
which, again, ignoring spatial dependencies and rescaling the
variables,  readily becomes the DRW equation, Eq.(\ref{1}).
The very same reasoning applies also to the other universality classes
discussed in the Introduction (i.e. dynamical percolation and the Manna
class); also in these cases the corresponding non-linear terms,
describing saturation effects vanish upon properly rescaling the system.

On the other hand, beyond the mean-field limit, the non-linearities
are essential and control the ``renormalized'' values of the avalanche
exponents (see e.g. \cite{Beyond}), which differ for the various
universality classes \cite{avalanches,Lubeck}, and avalanches can develop
non-symmetric shapes \cite{Mikko}.

\section{First-return time distributions}
Following the general result of A. Bray \cite{Bray} (see also
F. Colaiori \cite{Colaiori2}), here we summarize the computation of
avalanche exponents for a random walk in a logarithmic potential.
The general Fokker-Plank equation reads \cite{Gardiner}
\begin{equation}
\frac{\partial P(x,t)}{\partial t}=\mu\frac{\partial}{\partial
  x}\left(\frac{\partial P(x,t)}{\partial
    x}+\frac{\beta}{x}P(x,t)\right).
\label{fp}
\end{equation}
To calculate the probability distribution $F(T)$ of the return times
at which a walker starting close to the origin
($P(x,0)=\delta(x-\epsilon), \epsilon\rightarrow0$) first hits back
the origin, the absorbing boundary condition $P(0,t)=0$ needs to be
imposed.  Note that $F(T)$ is minus the probability flux at $0$,
 $F(T)=-j(0,t=T)$, with
\begin{equation}
j(0,t=T)=-\mu\left[\frac{\partial P(x,t)}{\partial x}+\frac{\beta}{x}P(x,t)\right]_{x=0}.
\end{equation}
One can try  a solution of the Eq.(\ref{fp}) of the form
$P(x,t)=r(x)\exp(-\mu k^{2}t)$ and note that the resulting
equation can be converted into a Bessel Equation with the change of
variable $r(x)=x^{\frac{1-\beta}{2}}R(x)$, \begin{equation}
x^{2}R''(x)+xR'(x)+\left(k^{2}x^{2}-\nu^{2}\right)R(x)=0,
\end{equation}
where $\nu=(1+\beta)/2$. The general solution of this last equation is
a linear combination of Bessel functions of the first kind of order
$\pm\nu$. Putting the pieces back together,
employing the orthogonality property of the Bessel functions, and
imposing the initial condition, leads to
\begin{eqnarray}
  P(x,t\mid\epsilon,0)&=&\left(\frac{x}{\epsilon}\right)^{1-\nu}\epsilon\int_{0}^{\infty}dk
  k [
  AJ_{\nu}(k\epsilon)J_{\nu}(kx)
    \nonumber \\
   &+& BJ_{-\nu}(k\epsilon)J_{-\nu}(kx) ]e^{-\mu k^{2}t},
\label{JJ}
\end{eqnarray}
where $A$ and $B$ are numerical constants. The integral in
Eq.(\ref{JJ}) gives the modified Bessel function of the first kind
$I_{\pm\nu}$ and, it is easy to compute the flux at the origin in the
small $\epsilon$ limit \cite{Bray,Colaiori2}, leading to
Eq.(\ref{return}).

\vspace{1cm}

\begin{acknowledgments}
  We are grateful to the Spanish-MINECO for financial support (under
  grant FIS2013-43201-P; FEDER funds). We warmly thank Francesca
  Colaiori, Jordi Hidalgo, and Paolo Moretti, for very useful comments
  and suggestions.
\end{acknowledgments}

%

\end{document}